\documentclass[a4paper,10pt]{article}
\usepackage {epsfig}
\begin{document}
\author{D. B. Ion$^{1)}$ and M. L. Ion$^{2)}$\\
$^{1)}${ National Institute for Physics and Nuclear Engineering,}\\
NIPNE-HH, Bucharest, P. O. Box MG-6, Romania\\
$^{2)}${ Bucharest University, Faculty of Physics, Department of Nuclear}%
\\
Physics, Bucharest, Romania}
\title{{\bf Nonextensive quantum statistics and saturation of the PMD-SQS
optimality limit in hadron-hadron scattering }}
\maketitle

\begin{abstract}
In this paper, new results on the analysis in hadron-hadron scattering ($\pi
N$, $KN$,  $ \overline{K}N$, etc) are obtained by using the {\it nonextensive quantum
entropy} and {\it principle of minimum distance in the space of quantum
states (PMD-SQS)}. So, using $[S_{J}(p)$, $S_{\theta }(q),S_{J\theta }(p),%
\overline{S}_{J\theta }(p,q)]$-{\it Tsallis-like scattering entropies}, {\it %
the optimality } as well as the {\it nonextensive statistical behavior of
the }[$J${\it \ and }$\theta $]{\it -quantum systems of states }produced in
hadronic scatterings are investigated in an unified manner. A connection
between optimal states obtained from the {\it principle of minimum distance
in the space of quantum states} (PMD-SQS) [17]\ and the most stringent
(MaxEnt) entropic bounds on Tsallis-like entropies for quantum scattering,
is established. The generalized entropic uncertainty relations as well as
the correlation between the nonextensivities p and q of the [$J${\it \ and }$%
\theta $]-statistics are proved. New results on the experimental tests of
the saturation of {\it the PMD-SQS-optimality limit}, as well as on the test
of {\it optimal entropic\ bands} obtained by using the experimental
pion-nucleon, kaon-nucleon, antikaon-nucleon phase shifts%
{\it , }are presented. The nonextensivity indices p and q are determined
from the experimental entropies by a fit with the optimal entropies $%
[S_{J}^{o1}(p)$, $S_{\theta }^{o1}(q),\overline{S}_{J\theta }^{o1}(p,q)]]$
obtained from the{\it \ principle of minimum distance in the space of states}.
In this way strong experimental evidences for the $p-$nonextensivities
index in the range $p=0.6$ with $q=p/(2p-1)=3,$ is obtained from the
experimental data of the ($\pi N,KN,\overline{K}N)$-scattering. The
experimental evidence obtained here for the {\it nonextensive statistical}
{\it behavior} {\it of the} $(J,\theta )-${\it quantum scatterings states}
in the above hadron-hadron scattering can be interpreted as an indirect
manifestation of the presence of the {\it quarks and gluons }as fundamental
constituents of the scattering system having the {\it strong-coupling
long-range regime required by the Quantum Chromodynamics}.
\end{abstract}

\section{Introduction}

In the last time there is an increasing interest in the foundation of a new
statistical theory [1,2] valid for the nonextensive statistical systems
which exhibit some relevant long range interactions, the memory effects or
multifractal structures. It is important to mention here that the Tsallis
nonextensive statistical formalism [2] already has been successfully applied
to a large variety of phenomena such as (see Ref. [3]): Levy-like  and
correlated anomalous diffusions, turbulence in electron
plasma, self-graviting systems, cosmology, galaxy
clusters, motion of Hydra viridissima, classical and quantum
chaos, quantum entanglement, reassociation in folded proteins,
superstatistics, economics, linguistic, etc. Here, is worth to
mention the recent applications of nonextensive statistics to nuclear and
high-energy particle physics, namely: electron-positron annihilations
[4,5], quark-qluon plasma [6], hadronic collisions [7-12,13],
nuclear collisions [14], and solar neutrinos [15,16].

In this paper, some new results on the optimal state analysis of
hadron-hadron scattering ($\pi N,KN,\overline{K}N,etc)$, obtained by using
the {\it nonextensive quantum entropy} [7-9] and {\it principle of minimum
distance in the space of quantum states (PMD-SQS) }[17], are
presented. Then, using $[S_{J}(p)$, $S_{\theta }(q)$, $S_{J\theta }(p)$, $\overline{S}_{J\theta }(p,q)]$-Tsallis-like scattering entropies, {\it %
the optimality } as well as the {\it nonextensive statistical behavior} of
the [$J$ and $\theta $]-quantum systems of states produced in
hadronic scatterings are investigated in an unified manner. A connection
between optimal states obtained from the {\it principle of minimum distance
in the space of quantum states} (PMD-SQS) [17]\ and the most stringent
(MaxEnt) entropic bounds on Tsallis-like entropies for quantum scattering,
is established. The nonextensivity indices p and q are determined from the
experimental entropies by a fit with the optimal entropies $[S_{J}^{o1}(p)$,
$S_{\theta }^{o1}(q),\overline{S}_{J\theta }^{o1}(p,q)]]$. In this way
strong experimental evidences for the $p-$nonextensivities index in the
range $p=0.6$ with $q=p/(2p-1)=3,$ are confirmed from the experimental data
of the principal hadron-hadron scatterings.

\section{\bf Optimality and nonextensive entropy for quantum scattering}

\subsection{\protect\smallskip {\bf Principle of minimum distance in the
space of quantum states}}

Recently in [17 ] we described the essential features of the hadron-hadron
scattering by using a new principle of optimum called{\it \ principle of
minimum distance in the space of quantum states }(PMD-SQS). Then knowledge
about the hadron-hadron scattering system (or more concretely, about partial
amplitudes) are deduced by assuming that the scattering system behaves as to
optimize some given measure of the system  effectiveness, e.g., {\it the
distance in the Hilbert space of scattering states.} Thus the behavior of
the scattering system is completely specified by those variational variables
(e.g., the partial scattering amplitudes) which are obtained by applying
constrained optimization to its effectiveness. The
PMD-SQS-optimum principle was formulated in a more general mathematical
form by using {\it reproducing kernel Hilbert spaces methods} [17-19].
Then, a new ''{\it analytic}'' quantum physics is developed in terms of the
reproducing functions from the reproducing kernel Hilbert spaces
(RKHS) of the transition amplitudes. In this new kind of {\it analytic
quantum physics} the system variational variables  are the partial
transition amplitudes which are introduced by the development of S-matrix
elements in terms of Fourier components, implied by the fundamental
symmetry of the quantum interacting system. Here  we discuss two very simple
cases, namely, the application of {\it PMD-SQS}$-${\it optimal} {\it %
principle} [17] to the $(\pi N,KN,\overline{K}N)-$scatterings.

Therefore, let $f_{++}(x)$ and $f_{+-}(x)$, $x\in [-1,1]$, be the scattering
helicity amplitudes of the meson -nucleon scattering process:

\begin{equation}
0^{-}+\frac{1}{2}^{+}\rightarrow 0^{-}+\frac{1}{2}^{+}  \label{1}
\end{equation}
$x=\cos (\theta ),\theta $ being the c.m. scattering angle. The
normalization of the helicity amplitudes $f_{++}(x)$ and $f_{+-}(x)$ is
chosen such that the c.m. differential cross section $\frac{d\sigma }{%
d\Omega }(x)$ is given by

\begin{equation}
\frac{d\sigma }{d\Omega }(x)=\mid f_{++}(x)\mid ^{2}+\mid f_{+-}(x)\mid ^{2}
\label{2}
\end{equation}

Since we will work at fixed energy, the dependence of $\sigma _{el}$ and $%
\frac{d\sigma }{d\Omega }(x)$ on this variable was
suppressed. Hence, the helicities of incoming and outgoing nucleons are
denoted by $\mu $, $\mu ^{^{\prime }}$, and was written as (+),(-),
corresponding to $(\frac{1}{2})$ and $(-\frac{1}{2})$, respectively. In
terms of the partial waves amplitudes $f_{J+}$ and $f_{J-}$ we have
\begin{equation}
\begin{tabular}{l}
$f_{++}(x)=\sum_{J=\frac{1}{2}}^{J_{\max }}(J+1/2)(f_{J-}+f_{J+})d_{\frac{1}{%
2}\frac{1}{2}}^{J}(x)$ \\
$f_{+-}(x)=\sum_{J=\frac{1}{2}}^{J_{\max }}(J+1/2)(f_{J-}-f_{J+})d_{-\frac{1%
}{2}\frac{1}{2}}^{J}(x)$%
\end{tabular}
\label{3}
\end{equation}
where the $d_{\mu \nu }^{J}(x)$-rotation functions are given by

\begin{equation}
d_{\frac{1}{2}\frac{1}{2}}^{J}(x)=\frac{1}{l+1}\cdot \left[ \frac{1+x}{2}%
\right] ^{\frac{1}{2}}\left[ \stackrel{\circ }{P}_{l+1}(x)-\stackrel{\circ }{%
P}_{l}(x)\right]  \label{4}
\end{equation}

\begin{equation}
d_{-\frac{1}{2}\frac{1}{2}}^{J}(x)=\frac{1}{l+1}\cdot \left[ \frac{1-x}{2}%
\right] ^{\frac{1}{2}}\left[ \stackrel{\circ }{P}_{l+1}(x)+\stackrel{\circ }{%
P}_{l}(x)\right]  \label{5}
\end{equation}
where $\stackrel{\circ }{P}_{l}$ are the derivatives of the Legendre polynomials.

Now, the elastic integrated cross section for the meson-nucleon
scattering can be expressed in terms of partial amplitudes $f_{J+}$ and $%
f_{J-}$
\begin{equation}
\sigma _{el}/2\pi =\int_{-1}^{+1}dx\frac{d\sigma }{d\Omega }(x)=\sum_{J=%
\frac{1}{2}}^{J_{\max }}(2J+1)(\mid f_{J+}\mid ^{2}+\mid f_{J-}\mid ^{2})
\label{6}
\end{equation}

Therefore, the variational variables for the $(0^{-}1/2^{+}\rightarrow
0^{-}1/2^{+})-$scatterings are the helicity amplitudes $f_{J+}$ and $f_{J-}$
while elastic integrated cross section expressed in terms of variational
variables by Eq. (6) is taken as the measure of system effectiveness.

Moreover, the elastic integrated cross section is directly related to the concept of quantum distance in the space of states. If H be the Hilbert space of the scattering states, defined
on the interval S$\equiv ${\bf $[-1,1]$}, with the inner product $<.$$%
\,,.>\, $ and the norm $\left\| \cdot \right\| ,$ given by
\[
\begin{tabular}{l}
$<f,g>=\int_{-1}^{+1}[f_{++}(x)\overline{g_{++}}(x)+f_{+-}(x)\overline{g_{+-}%
}(x)(x)]dx$ \\
$\sigma _{el}/2\pi =\int_{-1}^{+1}\frac{d\sigma }{d\Omega }%
(x)dx=\int_{-1}^{+1}[\mid f_{++}(x)|^{2}+\mid f_{+-}(x)\mid
^{2}]dx=||f||^{2} $%
\end{tabular}
\]
Then, the in general the distance $D(f,g)$ between any two scattering states
$f,g\in H$ is given by:
\[
D(f,g)=\min_{\Phi }\,\,\Vert \,f-g\exp (-i\Phi )\Vert =[\Vert f\Vert
^{2}+\Vert g\Vert ^{2}-2\mid <f,g>\mid ]^{\frac{1}{2}}
\]

The value of the arbitrary phase $\Phi _{\min }$ \thinspace for which the
distance function $\Vert f-g\exp (-i\Phi )\Vert $ is minimized is called
{\it minimum phase} (see e.g. Refs. [37-40]) and is given by : $\exp (i\Phi
_{\min })=<f,g>/\mid <f,g>\mid .$

If we take $g\equiv0$ then
\[
D(f,0)=\vert f\vert = \left[\frac{ \sigma_el}{2\pi }\right]^\frac{1}{2}
\]
As is seen from the above definition, the
{\it quantum distances }from the Hilbert space of the scattering amplitudes
have just the dimensions of length (e.g. fm, cm, etc.). The detalied
presentation of the PMD-SQS as well as some important results on application
of reproducing kernel Hilbert space (RKHS) methods to the extremal problems
of hadronic scattering can be found in Refs.[12,17,18]. Some definitions and
PMD-SQS-predictions are presented without a proof in the Tables 1-2.

\subsection{\bf J-nonextensive statistics for the quantum scattering\ states}

We define two kind of Tsallis{\it -}like scattering entropies. One of them,
namely $S_{J}(p),$ $p\in R,$ is special dedicated to the investigation of
the nonextensive statistical behavior of the {\it angular momentum} $J-${\it %
quantum states}, and can be defined by [7]

\begin{equation}
S_{J}(p)=\left[ 1-\sum (2J+1)p_{J}^{p}\right] /(p-1),\smallskip \ \;%
p\in R,\;  \label{7}
\end{equation}
where the probability distributions $p_{J}$ are given by

\begin{equation}
p_{J}=\frac{\mid f_{J+}\mid ^{2}+\mid f_{J-}\mid ^{2}}{\sum_{J=\frac{1}{2}%
}^{J_{\max }}(2J+1)(\mid f_{J+}\mid ^{2}+\mid f_{J-}\mid ^{2})},\smallskip \
\sum_{J=\frac{1}{2}}^{J_{\max }}(2J+1)\; p_{J}=1  \label{8}
\end{equation}
Here, it is important to present the following remark about {\it geometric
origin of the nonextensivity index p}$\in R.$

{\bf Remark 1:} Any Tsallis-like entropy of form (7) can be written in the
equivalent form

\[
\begin{tabular}{c}
$S_{J}(p)=\left[ 1- \left(|| \varphi _{J}||_{2p}\right)^{2p}\right]
/(p-1),\bigskip \ \smallskip $ \\
${\bf \varphi }_{J^{\pm }}=$ $f$ $_{J^{\pm }}/\sqrt{\sum_{J=\frac{1}{2}%
}^{J_{\max }}(2J+1)(\mid f_{J+}\mid ^{2}+\mid f_{J-}\mid ^{2})}$ \\
$with:\ |{\bf \varphi }_{J^{+}}|^{2}+\ |{\bf \varphi }_{J^{-}}|^{2}=p_{J},and
$ $\{{\bf \varphi }_{J}\}\in L_{2p},$ \\
$||{\bf \varphi }_{J}{\bf ||}_{2p}=\left[ \sum (2J+1)p_{J}^{p}\right] ^{%
\frac{1}{2p}}=\left[ 1+(1-p)S_{J}(p)\right] ^{\frac{1}{2p}}$%
\end{tabular}
\]
and consequently the nonextensivity index is  determined by the dimension $2p$
of the Hilbert space L$_{2p}$ of normalized partial helicity amplitudes $\{$ $f_{J^{\pm }}/\sqrt{\sigma _{el}/2\pi }\}.$

\subsection{$\theta ${\bf -nonextensive statistics for the quantum
scattering states}{\it \ }}

In similar way, for the $\theta -$scattering states considered as
statistical canonical ensemble, we can investigate their nonextensive
statistical behavior by using an angular Tsallis-like scattering entropy $%
S_{\theta }(q)$ defined as [7]

\begin{equation}
S_{\theta }(q)=\left[ 1-\int_{-1}^{+1}dx[P(x)]^{q}\right] /(q-1),\smallskip
\ q\in R\;  \label{9}
\end{equation}
where
\begin{equation}
P(x)=\frac{2\pi }{\sigma _{el}}\,\cdot \frac{d\sigma }{d\Omega }(x),\; %
\smallskip \ \smallskip \ \int_{-1}^{1}P(x)dx=1  \label{10}
\end{equation}
with $\frac{d\sigma }{d\Omega }(x)$ and $\sigma _{el}$ defined by
Eqs.(2)-(3) and (6).

{\bf Remark 2:} Any Tsallis-like entropy of form (9) can be written in the
equivalent form

\[
\begin{tabular}{c}
$S_{\theta }(q)=\left[ 1-\left(||{\bf \phi ||}_{2q}\right) ^{2q}\right] /(q-1),\smallskip $
\\
$\phi ^{++}(x)=f_{++}(x)/\sqrt{\int_{-1}^{+1}dx\left[
|f_{++}(x)|^{2}+|f_{+-}(x)|^{2}\right] }$ \\
$\phi ^{+-}(x)=f_{+-}(x)/\sqrt{\int_{-1}^{+1}dx\left[
|f_{++}(x)|^{2}+|f_{+-}(x)|^{2}\right] }$ \\
with:\ $|\phi ^{++}(x)|^{2}+|\phi ^{+-}(x)|^{2}=P(x),\rm{\;and\;}$ $\left( \phi
^{++},\phi ^{+-}\right) \in L_{2q},$ \\
$||{\bf \phi ||}_{2q}=\left[ \int_{-1}^{+1}dx[P(x)]^{q}\right] ^{\frac{1}{2q}%
}=\left[ 1+(1-q)S_{\theta }(q)\right] ^{\frac{1}{2q}}$%
\end{tabular}
\]
and consequently the nonextensivity index $q$ is strictly determined by the
dimension $2q$ of the Hilbert space L$_{2q}$ of the normalized helicity amplitudes $\{\phi ^{++},\phi ^{+-}\}.$

\subsection{${\bf [J\theta ]}-${\bf Tsallis-like scattering entropies}}

Also we can define the following generalized Tsallis-like combined entropy
[11,12]

\begin{equation}
\overline{S}_{J\theta }(p,q)=\left[ 1-\sum
(2J+1)p_{J}^{p}\int_{-1}^{+1}dx[P(x)]^{q}\right] /(p-1),\smallskip \ p\in R%
\; , \; q\in R\;  \label{11}
\end{equation}

The above Tsallis-like scattering entropies posses two important properties.
First, in the limit $k\rightarrow 1,k\equiv p,q,$ the Boltzmann-Gibs kind of
entropies is recovered:
\begin{equation}
\lim_{p\rightarrow 1}S_{J}(p)=S_{J}(1)=-\sum (2J+1)p_{J}\ln p_{J}
\label{12}
\end{equation}
\begin{equation}
\lim_{q\rightarrow 1}S_{\theta }(q)=S_{\theta
}(1)=-\int_{-1}^{+1}dxP(x)\ln P(x)  \label{13}
\end{equation}

Secondly, these entropies are called Tsallis-like scattering entropies,
having the {\it nonextensivity properties }in the sense that
\begin{equation}
S_{A+B}(k)=S_{A}(k)+S_{B}(k)+(1-k)S_{A}(k)S_{B}(k),\smallskip \ k=p,q\in \rm{\bf R}
\label{14}
\end{equation}
for any independent sub-systems $A,B$ $(p_{A+B}=p_{A}\cdot p_{B}).$ Hence,
each of the indices $p\neq 1$ or $q\neq 1$ from the definitions (7) and (9)
can be interpreted as measuring the degree {\it nonextensivity.}

{\bf Remark 3:} Any Tsallis-like entropy of form (11) can be written in the
following equivalent form
\[
\begin{tabular}{c}
$\overline{S}_{J\theta }(p,q)=\left[ 1-\left(||{\bf \varphi }_{J}{\bf ||}%
_{2p}\right)^{2p}\left(||{\bf \phi ||}_{2q}\right)^{2q}\right] /(p-1)$ \\
$\left(||{\bf \varphi }_{J}{\bf ||}_{2p}\right)^{2p}\left(||{\bf \phi ||}_{2q}\right)^{2q}=[1+(1-p)%
\overline{S}_{J\theta }(p,q)]$%
\end{tabular}
\]

\subsection{{\bf The equilibrium distributions for the
[J]- and [$\theta $]- systems of quantum scattering states}}

We next consider the maximum-entropy (MaxEnt) problem
\begin{equation}
\max \{S_{J}(p),S_{\theta }(q),S_{\theta J}(q),\overline{S}_{J\theta }(p,q)\}%
\rm{\; when \;}\sigma _{el}=\rm{fixed \; and \;}\frac{d\sigma }{d\Omega }(1)=%
\rm{fixed}  \label{15}
\end{equation}
as criterion for the determination of the ''equilibrium'' distributions
$p_{l}^{me}$ and $P^{me}(x)$ for the system of quantum states produced by
the $(0^{-}\frac{1}{2}\rightarrow 0^{-}\frac{1}{2})-$scattering. The {\it %
equilibrium distributions, }as well as the{\it optimal scattering
entropies }for the quantum scattering of the spineless particles were
obtained in Ref. [8-9]. For the{\it  J-quantum states, }in the spin $%
(0^{-}\frac{1}{2}\rightarrow 0^{-}\frac{1}{2})$ scattering case, these
distributions are given by:
\begin{equation}
\begin{tabular}{l}
$p_{J}^{me}=p_{J}^{o1}=\frac{1}{2K_{\frac{1}{2}\frac{1}{2}}(+1,+1)}=\frac{1}{%
(J_{o}+1)^{2}-1/4},\smallskip \ \rm{\; for \;}\frac{1}{2}\leq J\leq J_{o},%
\rm{\; and}$ \\
$p_{J}^{me}=0, \rm{\; for \;} J\geq J_{o}+1$%
\end{tabular}
\label{16}
\end{equation}
while, for the $\theta -${\it quantum states}, these distributions are as
follows

\begin{equation}
P^{me}(x)=P^{o\pm 1}(x)=\frac{\left[ K_{\frac{1}{2}\frac{1}{2}}(x,1)\right]
^{2}}{K_{\frac{1}{2}\frac{1}{2}}(1,1)}  \label{17}
\end{equation}
where $d_{\frac{1}{2}\frac{1}{2}}^{J}(x)$ are the d-spin rotation functions
(4)-(5) for the spin 1/2 particles, $\stackrel{\circ }{P}_{l}(x)$
are the derivatives of Legendre polynomials. The{\it \ reproducing kernel }[17-19] $K_{\frac{1}{2}\frac{1}{%
2}}(x,1)$ is given by
\begin{equation}
K_{\frac{1}{2}\frac{1}{2}}(x,1)=\frac{1}{2}\sum_{1/2}^{J_{o}}(2J+1)d_{\frac{1%
}{2}\frac{1}{2}}^{J}(x)  \label{18}
\end{equation}
while the {\it optimal angular momentum} $J_{o\; }$ is
\begin{equation}
(J_{o}+1)^{2}-1/4=2K_{\frac{1}{2}\frac{1}{2}}(1,1)=\frac{4\pi }{\sigma _{el}}%
\frac{d\sigma }{d\Omega }(1)  \label{19}
\end{equation}
We note that results similar to (16)-(19) can be obtained with the
constraint: $\frac{d\sigma }{d\Omega }(-1)=fixed$ instead of $\frac{d\sigma
}{d\Omega }(1)=fixed$

In the Table 2 we presented analytic formulas for both maximization problem
of form (15).

{\it Proof:} In this case solving the problem (15) via Lagrange multipliers
[20] we obtain that the singular solution $\lambda _{0}=0$ exists and is
just given by the $[S_{J}^{o1}(p),S_{\theta }^{o1}(q),S_{\theta J}^{o1}(q)]-$%
{\it optimal entropies} corresponding to the PMD-SQS-{\it optimal state}
(see Table 1). Indeed, the problem (15) is equivalent to the following
unconstrained extremization problem [20]:
\begin{equation}
\begin{tabular}{c}
$\pounds \equiv \lambda _{0}\left\{ S_{J}(p),S_{\theta }(q),S_{\theta J}(p),%
\overline{S}_{J\theta }(p,q)\right\} +\lambda _{1}\left\{ \sigma _{el}/4\pi
-\sum (2J+1)\left[ \mid f_{J-}\mid ^{2}+\mid f_{J+}\mid ^{2}\right] \right\}
$ \\
$+\lambda _{2}\left\{ \frac{d\sigma }{d\Omega }(1)-\left[ \sum (2J+1)%
{\rm \bf \;Re}%
(f_{J+}+f_{J-})\right] ^{2}-\left[ \sum (2J+1)%
{\rm \bf \;Im}%
(f_{J+}+f_{J-})\right] ^{2}\right\} \rightarrow extremum$%
\end{tabular}
\label{20}
\end{equation}

Hence, the solution of the problem (20) in the singular case [20] $\lambda
_{0}=0$ is reduced just to the solution of the {\it minimum constrained
distance in space of quantum states (PMD-SQS):}
\begin{equation}
\sum (2J+1)\left[ \mid f_{J-}\mid ^{2}+\mid f_{J+}\mid ^{2}\right] \rm{
when \;}\frac{d\sigma }{d\Omega }(+1)=\rm{\; is\; fixed}  \label{21}
\end{equation}
with the optimal state solution
\begin{equation}
f_{++}^{o+1}(x)=f_{++}(+1)\frac{K_{\frac{1}{2}\frac{1}{2}}(x,+1)}{K_{\frac{1%
}{2}\frac{1}{2}}(+1,+1)},\bigskip\ f_{+-}^{o+1}(x)=0  \label{22}
\end{equation}

Therefore, by a straightforward calculus we obtain that the solution of the
problem (20) is given by

\begin{equation}
S_{J}^{o1}(p)=\left[ 1-[2K_{\frac{1}{2}\frac{1}{2}}(1,1)]^{1-p}\right]
/(p-1),  \label{23}
\end{equation}

\begin{equation}
S_{\theta }^{o1}(q)=\left[ 1-\int_{-1}^{+1}dx\left( \frac{\left[ K_{\frac{1}{%
2}\frac{1}{2}}(x,1)\right] ^{2}}{K_{\frac{1}{2}\frac{1}{2}}(1,1)}\right)
^{q}\right] /(q-1),\;   \label{24}
\end{equation}

\begin{equation}
\overline{S}_{J\theta }^{o1}(p,q)=\left[ 1-\left( \frac{1}{2K_{\frac{1}{2}%
\frac{1}{2}}(1,1)}\right) ^{p-1}\int_{-1}^{+1}dx\left( \frac{\left[ K_{\frac{%
1}{2}\frac{1}{2}}(x,1)\right] ^{2}}{K_{\frac{1}{2}\frac{1}{2}}(1,1)}\right)
^{q}\right] /(p-1)  \label{25}
\end{equation}
for $p>0$, $q>0$0, where the {\it reproducing kernel} $K_{\frac{1}{2}\frac{1}{2}}(x,1)$ is
given by Eq.(18).

\subsection{{\bf \ Correlations between [J] and [$\theta$ ]-
nonextensive statistics}}

A natural but fundamental question was addressed in Refs. [9-11], namely,
what kind correlation (if it exists) is expected to be observed between the
nonextensivity indices$\ p${\it \ }and $q$ corresponding to the \ $(p,J)$%
-nonextensive statistics described by\ $S_{J}(p)$\ and $(q,\theta )-$%
nonextensive statistics described by $S_{\theta }(q)$? So, in general, an
answer at this question is difficult to give for all values of the
nonextensivities $p,q\in R.$ However, if the Fourier transform defined by
Eqs. (3) is considered a bounded map from the space $L_{2p}$ of the vector
valued functions $\{(f_{J^{+}},f_{J^{-}}),$ $J=\frac{1}{2},\frac{3}{2},...\}$
and to the space $L_{2p\; }$of the vector valued functions $%
(f_{++},f_{+-}),$ respectively, then, the answer was given as follows
[11,12].

{\bf Riesz }$(\frac{1}{2p}+\frac{1}{2q}=1)${\bf -correlation:} Let \quad p$%
\in R^{+}$ and q$\in R^{+}$ be defined as the index 2p and 2q of the
Hilbert spaces L$_{2p}$ and L$_{2q}$ of the vector valued functions $%
(f_{J^{+}},f_{J^{-}}),$ and $(f_{++},f_{+-})$, respectively. Then, {\it th}e%
{\it \ nonextensivity indices p and q corresponding to the} [$J]-${\it %
statistics} {\it and [}$\theta ]-${\it statistics, respectively, are
expected to be correlated via the Riesz-Thorin relation}

\begin{equation}
\frac{1}{2p}+\frac{1}{2q}=1,\; \smallskip \ \rm{or \; }q=p/(2p-1)
\label{26}
\end{equation}
and the norm M of the {\it Fourier} {\it transform} [Eq. (3)-(4)] is
expected to be bounded by
\begin{equation}
M\equiv \frac{||Tf||_{L_{2q}}}{||f||_{L_{2p}}}\leq 2^{\frac{p-1}{2p}%
},\smallskip \ \frac{1}{2}<p<1\rm{\; and \;}q\; =p/(2p-1)  \label{27}
\end{equation}

{\it Proof: }In our case it was show that the result given by Eq. (26)-(27)
is a direct consequence of the Riesz-Thorin interpolation theorem extended
to the{\it \ vector-valued functions}. Indeed, let T be the Fourier
transform defined by the helicity scattering amplitude (3) where the partial
amplitudes are expressed as follow
\begin{equation}
f_{J^{\pm }}=\frac{1}{2}\int_{-1}^{+1}\left[ f_{++}(x)d_{\frac{1}{2}\frac{1}{%
2}}^{J}(x)\pm f_{+-}(x)d_{-\frac{1}{2}\frac{1}{2}}^{J}(x)\right] dx
\label{28}
\end{equation}
Then, it was shown that:
\begin{equation}
\sup \{|f_{J^{+}}|^{2}+|f_{J-}|^{2}\}^{1/2}\leq \frac{1}{\sqrt{2}}%
\int_{-1}^{+1}\left[ |f_{++}(x)|^{2}+|f_{+-}(x)|^{2}\right] ^{1/2}dx
\label{29}
\end{equation}
and it was used the Parseval's formula
\begin{equation}
\sum (2J+1)[|f_{J^{+}}|^{2}+|f_{J-}|^{2}]=\int_{-1}^{+1}\left[
|f_{++}(x)|^{2}+|f_{+-}(x)|^{2}\right] dx  \label{30}
\end{equation}
since $\left[ |d_{\frac{1}{2}\frac{1}{2}}^{J}(x)|^{2}+|d_{-\frac{1}{2}\frac{1%
}{2}}^{J}(x)|^{2}\right] ^{1/2}\leq 2^{1/2}$. This means that we have $%
T:L_{1}\rightarrow L_{\infty }$ with the norm $M_{1}=2^{-1/2}$ and $%
T:L_{2}\rightarrow L_{2}$ with the norm $M_{2}=1.$ Then, using the {\it %
Riesz-Thorin interpolation theorem }for the vector-valued functions{\it \ }%
(see J. Berth in Ref. [21]) $T:L_{p^{^{\prime }}}\rightarrow L_{q^{^{\prime
}}}$ with the norm M with $(1/p^{^{\prime }})=(1-t)/1+t/2,$ $(1/q^{^{\prime
}})=(1-t)/\infty +t/2,$ and $0<t<1.$ Hence, eliminating the parameter t $%
[t=(1/2q)-(1/2p)+1]$ and using the relations $p^{^{\prime }}=2p$ and $%
q^{^{\prime }}=2q,$ we get not only the condition (26) but also
the {\it norm - estimate} (27) since according to Riesz-Thorin
theorem [21]{\it \ } $M\leq M_{1}^{1-t}M_{2}^{t}$ .

\section{{\bf Numerical results}}

Now, for a systematic experimental investigation of the saturation of the
{\it optimality limits in hadron-hadron scattering }is necessary to use the
formulas from the Table 1 and the available experimental phase-shifts
[22-24] to solve the following important problems:

\begin{itemize}
\item  To reconstruct the ''{\it experimental}'' pion-nucleon, kaon-nucleon
and antikaon-nucleon scattering amplitudes;

\item  To obtain numerical values of the experimental scattering entropies $%
S_{J}(q),$ $S_{\theta }(q)$ from the reconstructed amplitudes;

\item  To obtain the numerical values of the {\it optimal } $J_{o}=\left[
\frac{4\pi }{\sigma _{el}}\frac{d\sigma }{d\Omega }(1)+1/4\right] ^{1/2}-1,$
from experimental scattering amplitudes and then, to calculate the numerical
values for the PMD-SQS-optimal entropies $S_{J}^{o1}(q),$ $S_{\theta
}^{o1}(q);$

\item  To obtain numerical values for $\chi _{J}^{2}(p)$ or/and $\chi
_{\theta }^{2}(q)$-{\it test functions} given by
\begin{equation}
\chi _{X}^{2}(k)=\sum_{i=1}^{n_{\exp }}\left[ \frac{%
[S_{X}(k)]_{i}-[S_{X}^{o1}(k)]_{i}}{[\Delta S_{X}^{o1}]_{i}}\right]
^{2},\quad X\equiv J,\theta ;\;\; k\equiv p,q  \label{31}
\end{equation}
where
\begin{equation}
\Delta S_{X}^{o1}(k)=|[S_{X}^{o1}(k)]_{J_{o}+1}-[S_{X}^{o1}(k)]_{J_{o}-1}|
\label{32}
\end{equation}
are the values of the $PMD-SQS-${\it optimal entropies} $%
[S_{X}^{o1}(k)]_{J_{o}\pm 1}$ calculated with the optimal angular momenta $%
J_{o}\pm 1,$ respectively. Of course, this procedure is equivalent with
assumption of an error of $\Delta J_{o}=\pm 1$ in estimation of the
experimental values of the optimal angular momentum $J_{o}.$ The results
obtained in this way are presented in the Fig. 1-3 and Table 3.
\end{itemize}

\subsection{{\bf Nonextensivity index p for the statistics of $J$ %
-quantum states}}

For the investigation of this important problem we use the experimental {\it %
pion-nucleon} [22] and {\it kaon-nucleon} [23] as well as antikaon-nucleon
phase-shifts [24] for the calculation of : $[S_{J}(p)]_{i},$ $%
[S_{J}^{o1}(p)]_{i}$ and $[\Delta S_{X}^{o1}]_{i}$ (see also Tables 1). The
values of $(\Delta S_{X}^{o1})_{i}$ are calculated by assuming an error of $%
\Delta J_{o}=\pm 1$ in the estimation of the {\it optimal angular momentum }$%
J_{o}$ from the experimental data [see Eq. (32)]. Then, by using Eq. (31) we
can calculate the values of $\chi _{J}^{2}(p).$ The numerical results
obtained in this way for $\chi _{J}^{2}(p)/n_{D}$ for different
nonextensivities p in the interval $0.5\leq k\leq 7.00$ are presented in the
Table 3, respectively. Hence, the results from Table 3 allow us to conclude
that the statistics of the system of $J-$quantum states are {\it %
superextensitive} (superadditive) with values of the nonextensivity index $p$
in the interval $1/2\leq p\leq 0.6.$ This experimental discovery can be
compared with the recent results of Refs. [23-25] about the observed radial
density profiles in pure-electron plasmas in Penning traps, which are also
consistent with a value of the nonextensivity index around $p=1/2$.

\subsection{{\bf Nonextensivity index q for the statistics of [$\theta ]$%
-quantum states}}

In similar way, from the experimental {\it pion-nucleon} [20], {\it %
kaon-nucleon} [21] and {\it antikaon-nucleon } [22] phase-shifts, we
obtain the experimental values of: $[S_{\theta }(q)]_{i},$ $[S_{\theta
}^{o1}(q)]_{i}$ and $[\Delta S_{\theta }^{o1}(q)]_{i}$ and, consequently,
the experimental values of $\chi _{\theta }^{2}(q)/n_{D}$ presented in Table
3 for the $[(\pi N)_{I=1/2,3/2};(KN)_{I=0,1};(\overline{K}N)_{I=0,1}]-$%
scatterings. From the results of the Table 3 we conclude that the statistics
of the system composed from $\theta -$quantum states are {\it subextensive}
(subadditive) with an index $q$ $\geq 3.$

\subsection{\bf Experimental evidence for (1/2p+1/2q=1)-nonextensivity
correlation}

Now, we can give an ''{\it experimental}'' answer to the fundamental
question: what kind of correlation (if it exists) is expected to be observed
between the nonextensivity indices$\ p${\it \ }and $q$ corresponding to the
\ $(p,J)$-nonextensive statistics described by\ $S_{J}(p)$\ and $(q,\theta
)- $nonextensive statistics described by $S_{\theta }(q)$? [We remember that
the ''{\it mathematical}'' answer is given by Eq. (26)]. Indeed, from
Figs.1-3 as well as from the Table 3 we see that the experimental data on
the scattering entropies $S_{J}(p)$ and $S_{\theta }(q)$ are simultaneously
in excellent agreement $(CL>99\%)$ with the $[S_{J}^{o1}(p)$, $S_{\theta
}^{o1}(q)]-$optimal state predictions if the nonextensivities p and q of the
$(J$ and $\theta )-$statistics are correlated via Riesz-Thorin
relation: $1/p+1/q=2$ (or $q=p/(2p-1)$). So, the best fit is obtained
(see Tables 3) for the correlated pairs $p$ and $q=p/(2p-1)$ with the values
of $p$ in the range $p=$ $0.6$ and $q=p/(2p-1)=3$.

\section{\bf Conclusions}

In this paper, by introducing $[S_{J}(p)$, $S_{\theta }(q),\overline{S}%
_{J\theta }(p,q)]$-Tsallis-like entropies, {\it the saturation of the
optimality limits } as well as the {\it nonextensive statistical behavior}
of the [$J$ and $\theta $]-quantum states produced in hadronic
scatterings are investigated in an unified manner for the pure isospin [$\pi
N\rightarrow (\pi N)_{I=1/2,3/2}$; $KN\rightarrow (KN)_{I=0,1}$; $\overline{K}%
N\rightarrow (\overline{K}N)_{I=0,1}$]-scattering states. The main results
and conclusions can be summarized as follows:

\begin{itemize}
\item  Using the available experimental phase shifts analysis we
calculated the numerical values for the [$S_{J}(p)$, $S_{\theta }(q)$, %
$\overline{S}_{J\theta }(p,q)$]-{\it Tsallis-like scattering entropies }for
the pure isospin $I-$scattering states: [$(\pi N)_{I=1/2,3/2}$; $(KN)_{I=0,1}$; $(%
\overline{K}N)_{I=0,1}$];

\item  We presented strong experimental evidence for the {\it saturation of
the} $[S_{J}^{o1}(p)$, $S_{\theta }^{o1}(q)$, $\overline{S}_{\theta J}^{o1}(q)]-$%
{\it PMD-SQS optimal limits} for all nonextensive ($J$, $\theta$ ,%
$\overline{J\theta }$)-statistical ensembles of quantum states %
produced in hadron-hadron scattering (see Figs.1-3 and Table 3). These
results allow to conclude that the $[J]$-{\it quantum} {\it system} and $%
[\theta ]$-{\it quantum} {\it system} are produced at ''{\it equilibrium}''
but with the [$\frac{1}{2p}+\frac{1}{2q}]-$ conjugated nonextensivities
p=0.6 and $q=p/(2p-1)=3$ in all investigated isospin scattering
states: [$(\pi N)_{I=1/2,3/2}$; $(KN)_{I=0,1}$; $(\overline{K}N)_{I=0,1}]$. So the
''geometric origin'' of the nonextensivities p and q (as dimensions of the
Hilbert spaces $L_{2p}$ and $L_{2q})$ as well as their correlations are
experimentally confirmed with high accuracy ($CL >99\%$);

\item  The strong experimental evidence obtained here for the {\it %
nonextensive statistical behavior} of the $(J,\theta )-$ %
quantum scatterings states in the pion-nucleon, kaon-nucleon and
antikaon-nucleon scatterings
can be interpreted as an indirect manifestation the presence of the {\it %
quarks and gluons }as fundamental constituents of the scattering system
having the {\it strong-coupling long-range regime} required by the Quantum
Chromodynamics.
\end{itemize}

Finally, we note that further investigations are needed since this {\it %
saturation of optimality limits} as well as {\it nonextensive statistical
behavior} of the quantum scattering, emphasized here with high
accuracy $(CL>99\%)$, can be a signature of a {\it new universal
law }of the quantum scattering.\

\smallskip \ \newpage\

\begin{center}
{\bf Table 1 :} The optimal distributions, reproducing kernels, optimal
entropies, entropic

bands, for the $(0^{-}1/2^{+}\rightarrow 0^{-}1/2^{+})-$scattering

\begin{tabular}[t]{|c|c|c|c|}
\hline
Nr. & Name & $(0^{-}1/2^{+}\rightarrow 0^{-}1/2^{+})-$scattering & See Ref.
\\ \hline
1 & Optimal inequalities & $\frac{d\sigma }{d\Omega }(x)\leq K_{\frac{1}{2}%
\pm \frac{1}{2}}(\pm 1,\pm 1)\parallel f\parallel ^{2}$ & [7] \\ \hline
2 & Optimal states & $f^{o+1}(x)=f(1)\frac{K_{\frac{1}{2}\frac{1}{2}}(x,+1)}{%
K_{\frac{1}{2}\frac{1}{2}}(1,1)},$\smallskip \ $f^{o+1}(x)=f(-1)\frac{K_{-%
\frac{1}{2}\frac{1}{2}}(x,-1)}{K_{-\frac{1}{2}\frac{1}{2}}(-1,-1)}$ & [11]
\\ \hline
3 &
\begin{tabular}{c}
Reproducing kernels \\
$K_{\frac{1}{2}\pm \frac{1}{2}}(x,\pm 1)$
\end{tabular}
 &
$\begin{array}{c}
K_{\frac{1}{2}\frac{1}{2}}(x,+1)=\sum_{\frac{1}{2}}^{J_{o}}(J+\frac{1}{2}%
)d_{\frac{1}{2}\frac{1}{2}}^{J}(x) \\
K_{\frac{1}{2}-\frac{1}{2}}(x,-1)=\sum_{\frac{1}{2}}^{J_{o}}(J+\frac{1}{2}%
)d_{\frac{1}{2}-\frac{1}{2}}^{J}(x) \\
2K_{\frac{1}{2}\pm \frac{1}{2}}(\pm 1,\pm 1)=(J_{o}+1)^{2}-1/4%
\end{array} $
 & [9-12,17] \\ \hline
4 &
\begin{tabular}{c}
Optimal distributions \\
$P^{o\pm 1}(x)$
\end{tabular}

 & $P^{o1}(x)=\frac{\left[ K_{\frac{1}{2}\frac{1}{2}}(x,1)\right] ^{2}}{K_{%
\frac{1}{2}\frac{1}{2}}(1,1)},\smallskip \ P^{o-1}(x)=\frac{\left[ K_{\frac{1%
}{2}-\frac{1}{2}}(x,-1)\right] ^{2}}{K_{\frac{1}{2}-\frac{1}{2}}(-1,-1)}$ &
[7-9] \\ \hline
4 &
\begin{tabular}{c}
Optimal distribution \\
$\{p_{J}^{o1}\}$
\end{tabular}
& \begin{tabular}{c}
$p_{J}^{o\pm 1}=\frac{1}{2K_{\frac{1}{2}\pm \frac{1}{2}}(\pm 1,\pm 1)},$ for
$1/2\leq J\leq J_{o}$ \\
$p_{J}^{o\pm 1}=0$ for $J\geq J_{o}+1$%
\end{tabular}
& [7-12] \\ \hline
\multicolumn{4}{|c|}{{*}*} \\ \hline
5 & $
\begin{array}{c}
\rm{Number \; of} \\
\rm{optimal \; states} \\
\rm{\; for \; y=1}
\end{array}
$ & $N_{o}=\sum (2J+1)=(J_{o}+1)^{2}-1/4=2K_{\frac{1}{2}\pm \frac{1}{2}}(\pm
1,\pm 1)$ & [7-8] \\ \hline
6 & $
\begin{array}{c}
\rm{Optimal} \\
\rm{angular-momentum}
\end{array}
$ & $J_{o}=\left\{ \left[ \frac{4\pi }{\sigma _{el}}\frac{d\sigma }{d\Omega }%
(\pm 1)+\frac{1}{4}\right] ^{1/2}-1\right\} $ & [7-8] \\ \hline
7 & $
\begin{array}{c}
\rm{Optimal \; entropy} \\
S_{L}^{o1}(q)
\end{array}
$ & $S_{L}^{o\pm 1}(p)=\frac{1}{p-1}\left[ 1-N_{os}^{1-p}\right] $ & [7-8]
\\ \hline
8 & $
\begin{array}{c}
\rm{Optimal \; entropy} \\
S_{\theta }^{o1}(q)
\end{array}
$ & $S_{\theta }^{o\pm 1}(q)=\frac{1}{q-1}\left[ 1-\int_{-1}^{+1}dx\left(
\frac{\left[ K_{\frac{1}{2}\pm \frac{1}{2}}(x,\pm 1)\right] ^{2}}{K(\pm
1,\pm 1)}\right) ^{q}\right] $ & [9-12] \\ \hline
9 & $
\begin{array}{c}
\rm{Optimal \; entropy} \\
S_{\theta L}^{o1}(q)
\end{array}
$ & $S_{\theta L}^{o\pm 1}(q)=\frac{1}{q-1}\left[
1-N_{o}^{1-q}\int_{-1}^{+1}dx\left( \frac{\left[ K_{\frac{1}{2}\pm \frac{1}{2%
}}(x,\pm 1)\right] ^{2}}{K_{\frac{1}{2}\pm \frac{1}{2}}(\pm 1,\pm 1)}\right)
^{q}\right] $ & [9-12] \\ \hline
10 & $
\begin{array}{c}
\rm{Optimal \; entropy} \\
\overline{S}_{\theta L}^{o1}(p)
\end{array}
$ & $\overline{S}_{\theta L}^{o\pm 1}(q)=\frac{1}{p-1}\left[
1-N_{o}^{1-p}\int_{-1}^{+1}dx\left( \frac{\left[ K_{\frac{1}{2}\pm \frac{1}{2%
}}(x,\pm 1)\right] ^{2}}{K_{\frac{1}{2}\pm \frac{1}{2}}(\pm 1,\pm 1)}\right)
^{q}\right] $ & [9-12] \\ \hline
11 & $J-$entropic band & $0\leq S_{J}(p)\leq S_{J}^{o\pm 1}(p)$ & [9-12] \\
\hline
12 & $\theta -$entropic band & $\frac{1}{q-1}[1-K(1,1)^{q-1}]\leq S_{\theta
}(q)\leq S_{\theta }^{o\pm 1}(q)$ & [9-12] \\ \hline
13 & $J\theta -$entropic band $(q=p)$ & $\frac{1}{q-1}\left[
1-2^{1-p}\right] \leq S_{\theta L}(q)\leq S_{\theta L}^{o\pm 1}(q)$ & [9-12]
\\ \hline
14 & $\overline{J\theta }-$entropic band $(q\neq p)$ & $\frac{1-q}{1-p}%
S_{\theta }^{o\pm 1}(q)\leq \overline{S}_{\theta L}(p)\leq \overline{S}%
_{\theta L}^{o\pm 1}(p)$ & [11,12] \\ \hline
\end{tabular}

\newpage

{\bf Table 2 :}{\large \ }Examples of optimal angular distributions $P_{o\pm
1}(x),$ optimal logarithmic slope, optimal scaling variable and optimal scaling function

\smallskip
\begin{tabular}{|c|c|}
\hline
& $(0^{-}1/2^{+}\rightarrow 0^{-}1/2^{+})-${\bf scatterings} \\ \hline
${\bf J}_{o}$ & $P^{o\pm 1}(x)=\frac{\left[ K_{\frac{1}{2}\pm \frac{1}{2}%
}(x,\pm 1)\right] ^{2}}{K_{\frac{1}{2}\pm \frac{1}{2}}(\pm 1,\pm 1)}$ \\
\hline
{\bf 1/2} & $\frac{1}{2}(1\pm x)$ \\ \hline
{\bf 3/2} & $\frac{3}{2}(1\pm x)x^{2}$ \\ \hline
{\bf 5/2} & $\frac{9}{48}(1\pm x)(5x^{2}-1)^{2}$ \\ \hline
{\bf 7/2} & $\frac{10}{32}(1\pm x)(7x^{2}-3)^{2}x^{2}$ \\ \hline
{\bf 9/2} & $\frac{15}{128}(1\pm x)(21x^{4}-14x^{2}+1)^{2}$ \\ \hline
{\bf 11/2} & $\frac{441}{10750}(1\pm x)(66x^{4}-60x^{2}+10)^{2}x^{2}$ \\
\hline
{\bf 13/2} & $\frac{1}{14336}(1\pm x)(3003x^{6}-3465x^{4}+945x^{2}-35)^{2}$
\\ \hline
{\bf 15/2} & $\frac{1}{1179648}(1\pm
x)(51480x^{6}-72072x^{4}+27720x^{2}-2520)^{2}x^{2}$ \\ \hline
& ** \\ \hline
1 & {\bf Optimal logarithmic slope}:{\large \ }$b_{o\pm 1}=\frac{\overline{%
\lambda }^{2}}{4}\left[ \frac{4\pi }{\sigma _{el}}\frac{d\sigma }{d\Omega }%
(\pm 1)-1\right] $ \\ \hline
{2} & {\bf Optimal scaling variable} :{\bf \ }$\tau _{o\pm 1}\equiv 2\left[
\left| t\right| b_{o\pm 1}\right] ^{1/2}=\left\{ \overline{\lambda }%
^{2}\left| t\right| \left[ \frac{4\pi }{\sigma _{el}}\frac{d\sigma }{d\Omega
}(\pm 1)-1\right] \right\} ^{1/2}$ \\ \hline
{3} & {\bf Optimal scaling }: $\frac{1}{\frac{d\sigma }{d\Omega }(\pm 1)}%
\frac{d\sigma ^{o\pm 1}}{d\Omega }{\bf (}x{\bf )}=\frac{P(x)}{P(\pm 1)}%
\simeq \left[ \frac{2J_{1}(\tau _{o\pm })}{\tau _{o\pm }}\right] ^{2},$ $%
J_{1}(\tau _{o\pm })-
\begin{tabular}{l}
{\it Bessel function} \\
{\it of first order}
\end{tabular}
$ \\ \hline
4 & {\bf Optimal inequality} :{\large \ }$b\equiv \frac{d}{dt}\ln \left[
\frac{d\sigma }{d\Omega }(s,t)\right] _{\mid t=0}\geq \frac{\overline{%
\lambda }^{2}}{4}\left[ \frac{4\pi }{\sigma _{el}}\frac{d\sigma }{d\Omega }%
(\pm 1)-1\right] =b_{o\pm 1}$ \\ \hline
\end{tabular}
\

***

\bigskip {\bf Table 3:} $\chi ^{2}/n_{D}$ obtained from comparisons of the
experimental scattering entropies: S$_{J}(p),$ S$_{\theta }(q),$ S$_{J\theta
}(p),$and $\overline{S}_{J\theta }(p,q),$ with the optimal entropies: S$%
_{J}^{o1}(p),$ S$_{\theta }^{o1}(q),$ S$_{J\theta }^{o1}(p),$and $\overline{S%
}_{J\theta }^{o1}(p,q),$ respectively, for the ($\pi N$ , $KN$, $\overline{K}%
N)$-scattering (see the text)

\begin{tabular}{||c||c||c||}
\hline\hline
\begin{tabular}{c}
Hadron-hadron \\
scattering
\end{tabular}
&
\begin{tabular}{cc}
p & q=$\frac{p}{2p-1}$%
\end{tabular}
&
\begin{tabular}{c}
$\chi ^{2}/n_{D}$ \\
\begin{tabular}{cccc}
S$_{J}(p)$ & S$_{\theta }(q)$ & S$_{J\theta }(p)$ & $\overline{S}_{J\theta
}(p,q)$%
\end{tabular}
\end{tabular}
\\ \hline\hline
$
\begin{tabular}{c}
$\pi N\rightarrow (\pi N)_{I=1/2}$ \\
$88$ $PSA$ \\
$P_{LAB}$=$0.02\div 10$ $GeV/c$%
\end{tabular}
$ &
\begin{tabular}{c}
0.6 \\
1.0 \\
3.0
\end{tabular}
\begin{tabular}{c}
3.0 \\
1.0 \\
0.6
\end{tabular}
&
\begin{tabular}{cccc}
0.102 & 0.015 & 8.965 & 0.054 \\
0.649 & 0.648 & 124.2 & 0.721 \\
143.6 & 2.181 & 1. 10$^{6}$ & 312.6
\end{tabular}
\\ \hline\hline
\begin{tabular}{c}
$\pi N\rightarrow (\pi N)_{I=3/2}$ \\
$88$ $PSA$ \\
$P_{LAB}=0.02\div 10$ $GeV/c$%
\end{tabular}
&
\begin{tabular}{c}
0.6 \\
1.0 \\
3.0
\end{tabular}
\begin{tabular}{c}
3.0 \\
1.0 \\
0.6
\end{tabular}
&
\begin{tabular}{cccc}
0.130 & 0.090 & 8.456 & 0.105 \\
0.691 & 1.059 & 89.44 & 0.147 \\
209.2 & 3.010 & 8 10$^{5}$ & 63.15
\end{tabular}
\\ \hline\hline
\begin{tabular}{c}
$KN\rightarrow (KN)_{I=0}$ \\
$52$ $PSA$ \\
$P_{LAB}=0.1\div 2.65$ $GeV/c$%
\end{tabular}
&
\begin{tabular}{c}
0.6 \\
1.0 \\
3.0
\end{tabular}
\begin{tabular}{c}
3.0 \\
1.0 \\
0.6
\end{tabular}
&
\begin{tabular}{cccc}
0.449 & 0.035 & 13.55 & 0.014 \\
0.146 & 0.494 & 33.12 & 0.068 \\
0.190 & 1.089 & 145.8 & 0.240
\end{tabular}
\\ \hline\hline
\begin{tabular}{c}
$KN\rightarrow (KN)_{I=1}$ \\
$53$ $PSA$ \\
$P_{LAB}=0.05\div 2.65$ $GeV/c$%
\end{tabular}
&
\begin{tabular}{c}
0.6 \\
1.0 \\
3.0
\end{tabular}
\begin{tabular}{c}
3.0 \\
1.0 \\
0.6
\end{tabular}
&
\begin{tabular}{cccc}
0.089 & 0.045 & 1.567 & 0.011 \\
0.259 & 0.586 & 0.485 & 0.050 \\
30.53 & 1.030 & 2160. & 3.113
\end{tabular}
\\ \hline\hline
\begin{tabular}{c}
$\overline{K}N\rightarrow (\overline{K}N)_{I=0}$ \\
$50$ $PSA$ \\
$P_{LAB}=0.36\div 1.34$ $GeV/c$%
\end{tabular}
&
\begin{tabular}{c}
0.6 \\
1.0 \\
3.0
\end{tabular}
\begin{tabular}{c}
3.0 \\
1.0 \\
0.6
\end{tabular}
&
\begin{tabular}{cccc}
0.168 & 0.009 & 5.267 & 0.026 \\
0.199 & 0.267 & 9.084 & 0.028 \\
11.38 & 0.551 & 31.56 & 0.150
\end{tabular}
\\ \hline\hline
\begin{tabular}{c}
$\overline{K}N\rightarrow (\overline{K}N)_{I=1}$ \\
$50$ $PSA$ \\
$P_{LAB}=0.36\div 1.34$ $GeV/c$%
\end{tabular}
&
\begin{tabular}{c}
0.6 \\
1.0 \\
3.0
\end{tabular}
\begin{tabular}{c}
3.0 \\
1.0 \\
0.6
\end{tabular}
&
\begin{tabular}{cccc}
0.062 & 0.010 & 4.254 & 0.001 \\
0.064 & 0.196 & 5.065 & 0.007 \\
16.00 & 0.391 & 31.40 & 0.787
\end{tabular}
\\ \hline\hline
\end{tabular}
\end{center}
\newpage
\noindent
\begin{center}
\epsfig{width = 90mm, file=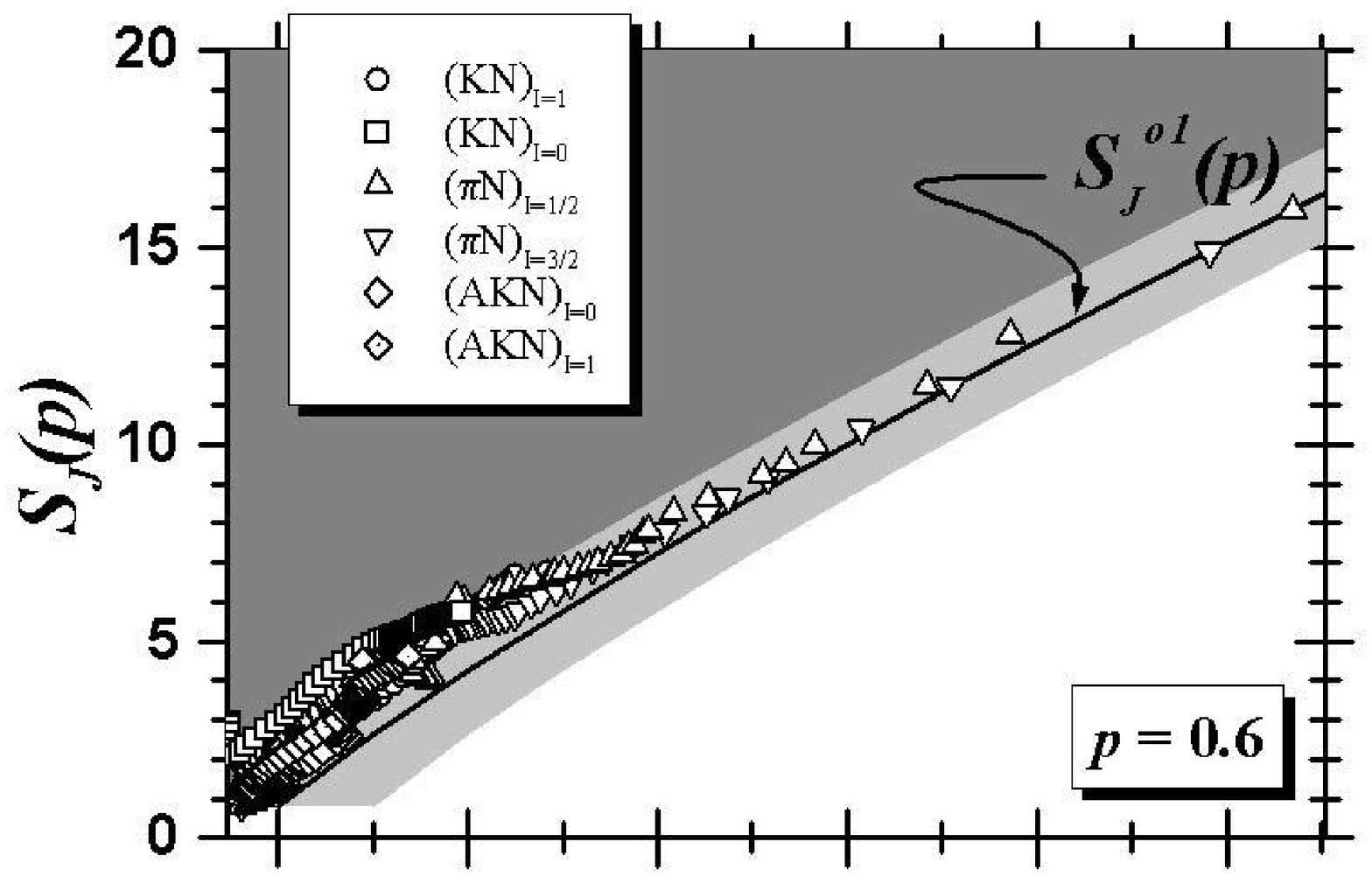}\\
\epsfig{width = 90mm, file=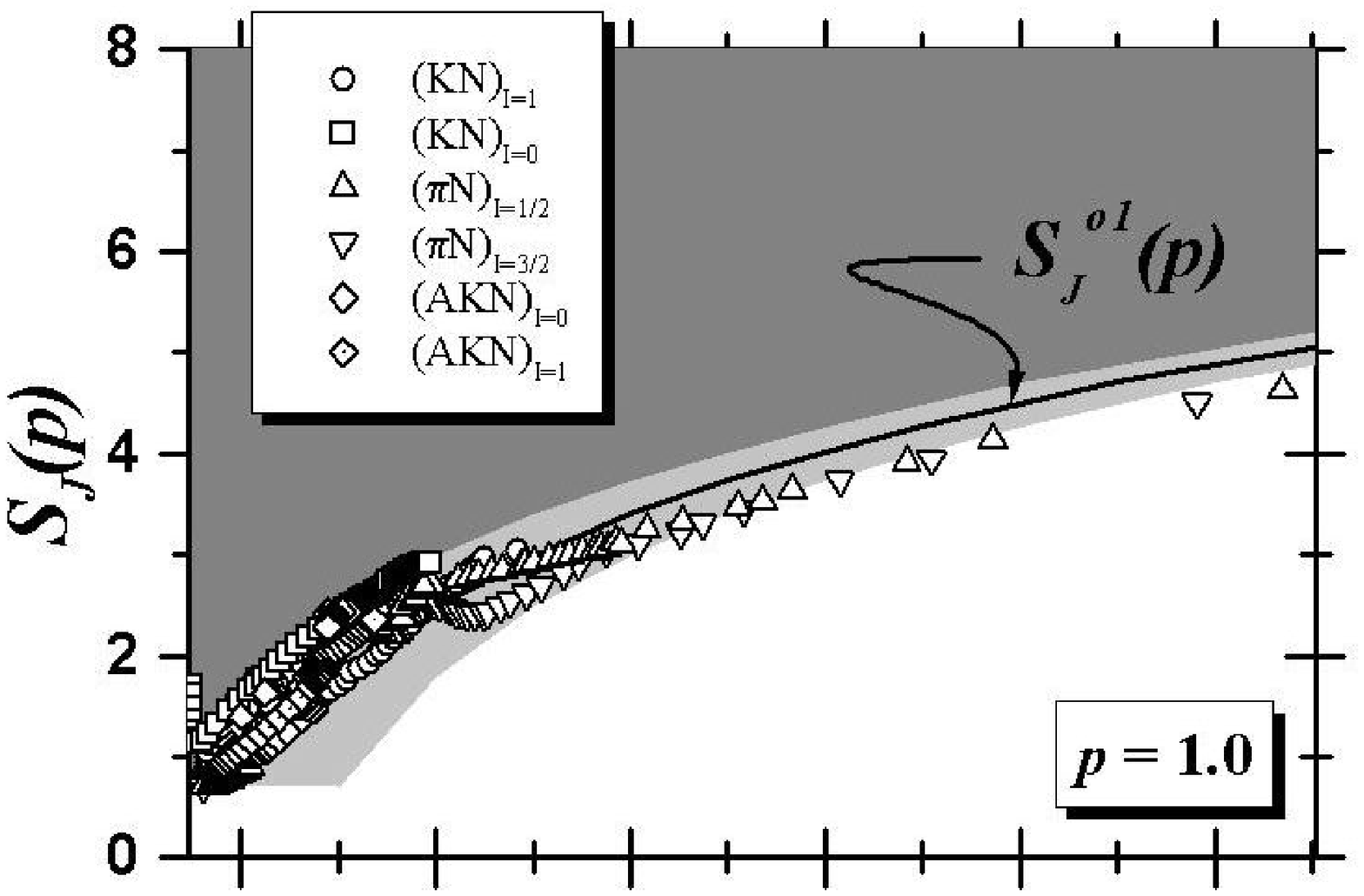}\\
\epsfig{width = 90mm, file=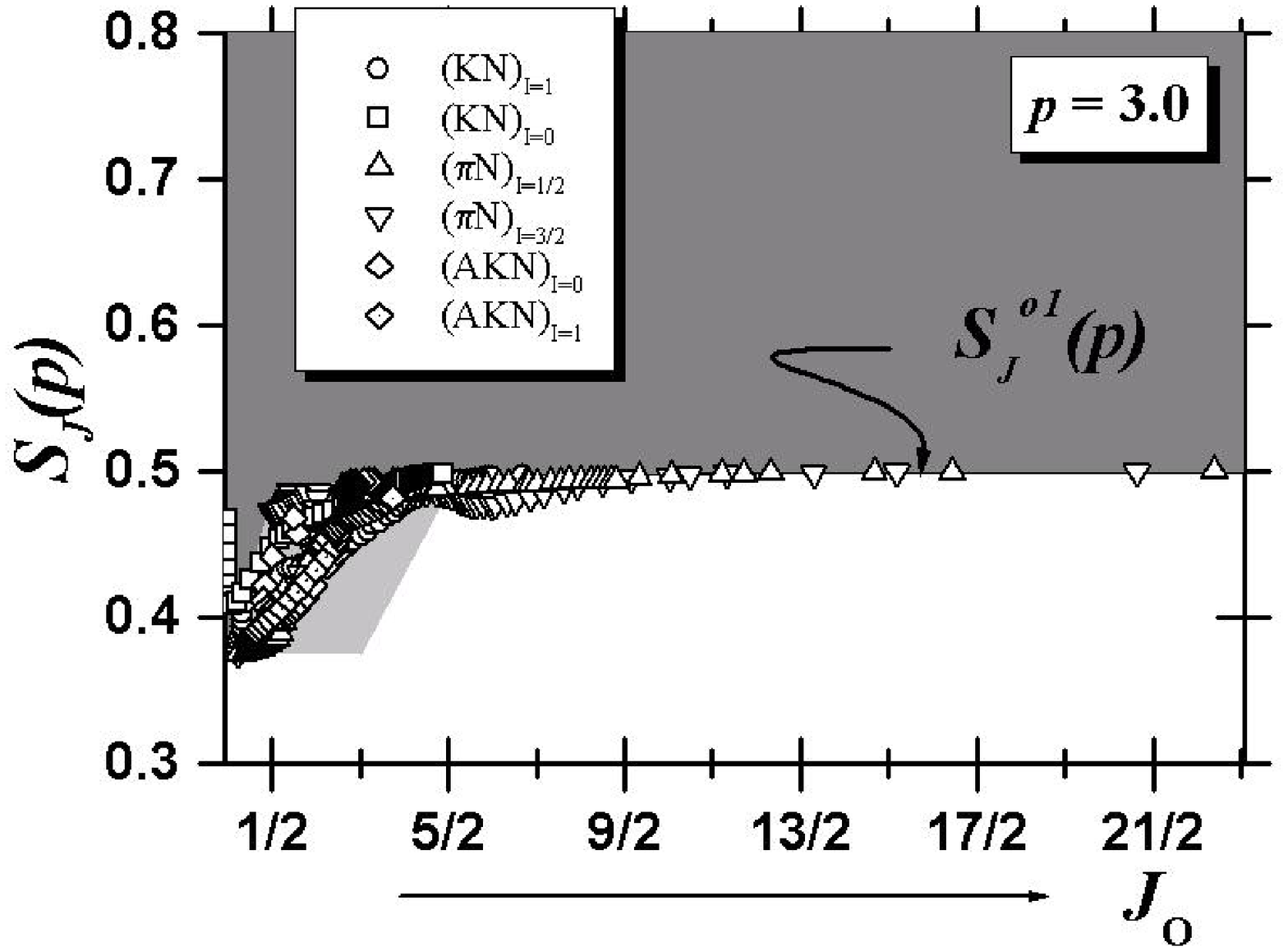}\\
\end{center}
{\bf Fig. 1: }  The experimental values of the Tsallis-like entropies
$S_{J}(p)$ for $[(\pi N)_{I=1/2,3/2};(KN)_{I=0,1};(\overline{K}N)_{I=0,1}]-$
scatterings, obtained from the available experimental phase-shifts [22-24],
are compared with the PMD-SQS-optimal state predictions $S_{J}^{o1}(p)$
given in Table 1 (full curve). The saturation of the PMD-SQS (MaxEnt)
optimal limits is evident.

\newpage

\noindent
\begin{center}
\epsfig{width = 90mm, file=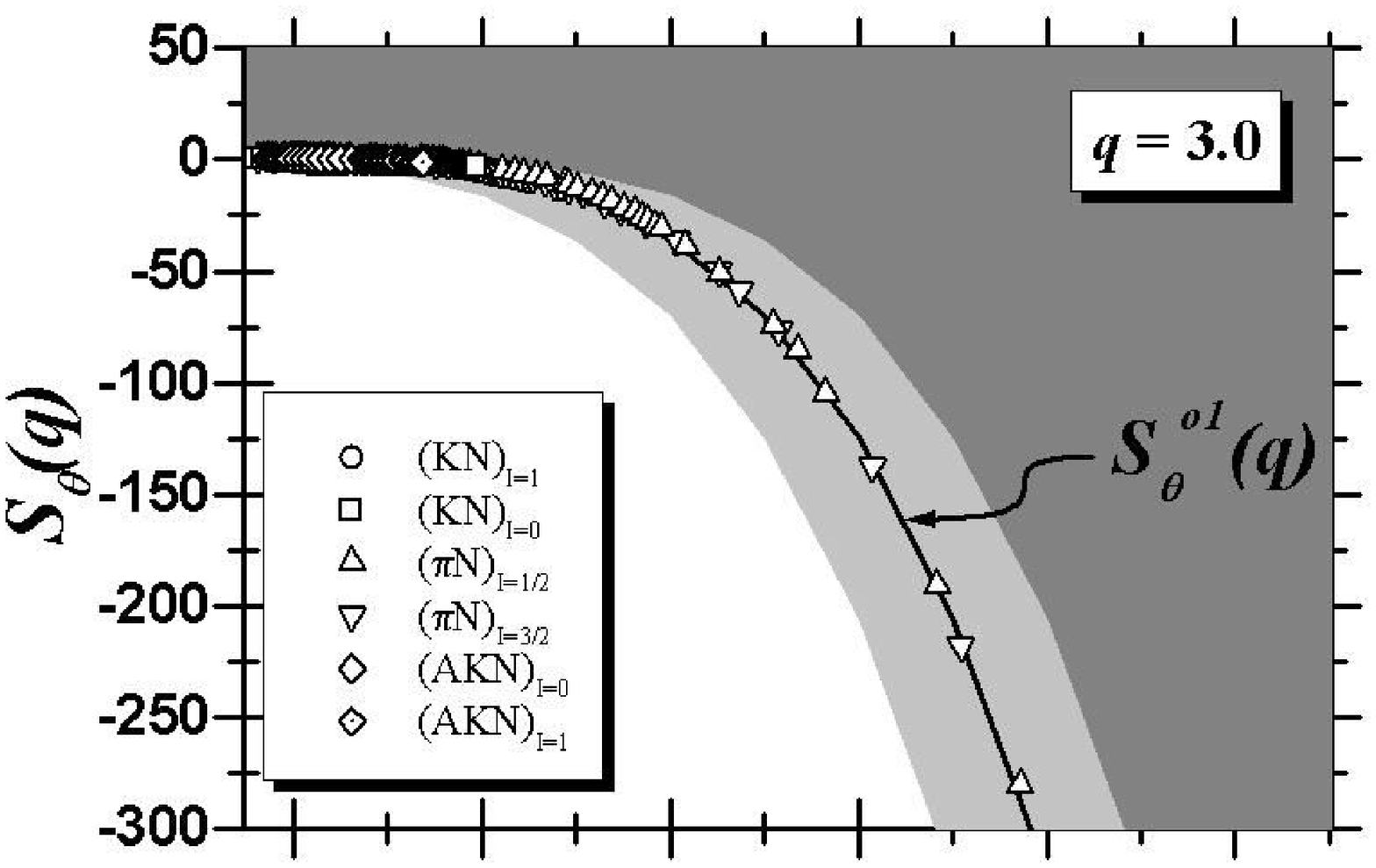}\\
\epsfig{width = 90mm, file=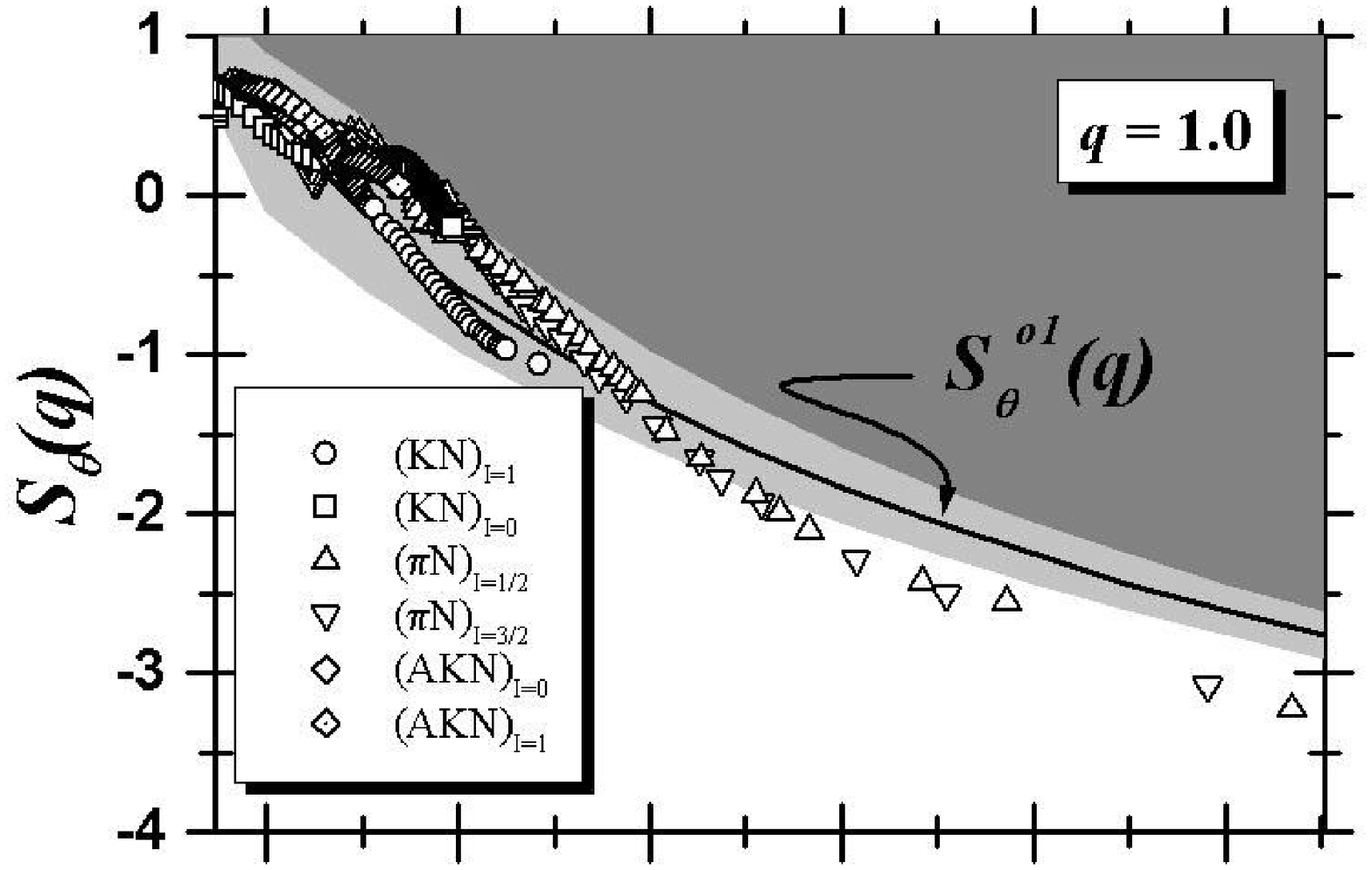}\\
\epsfig{width = 90mm, file=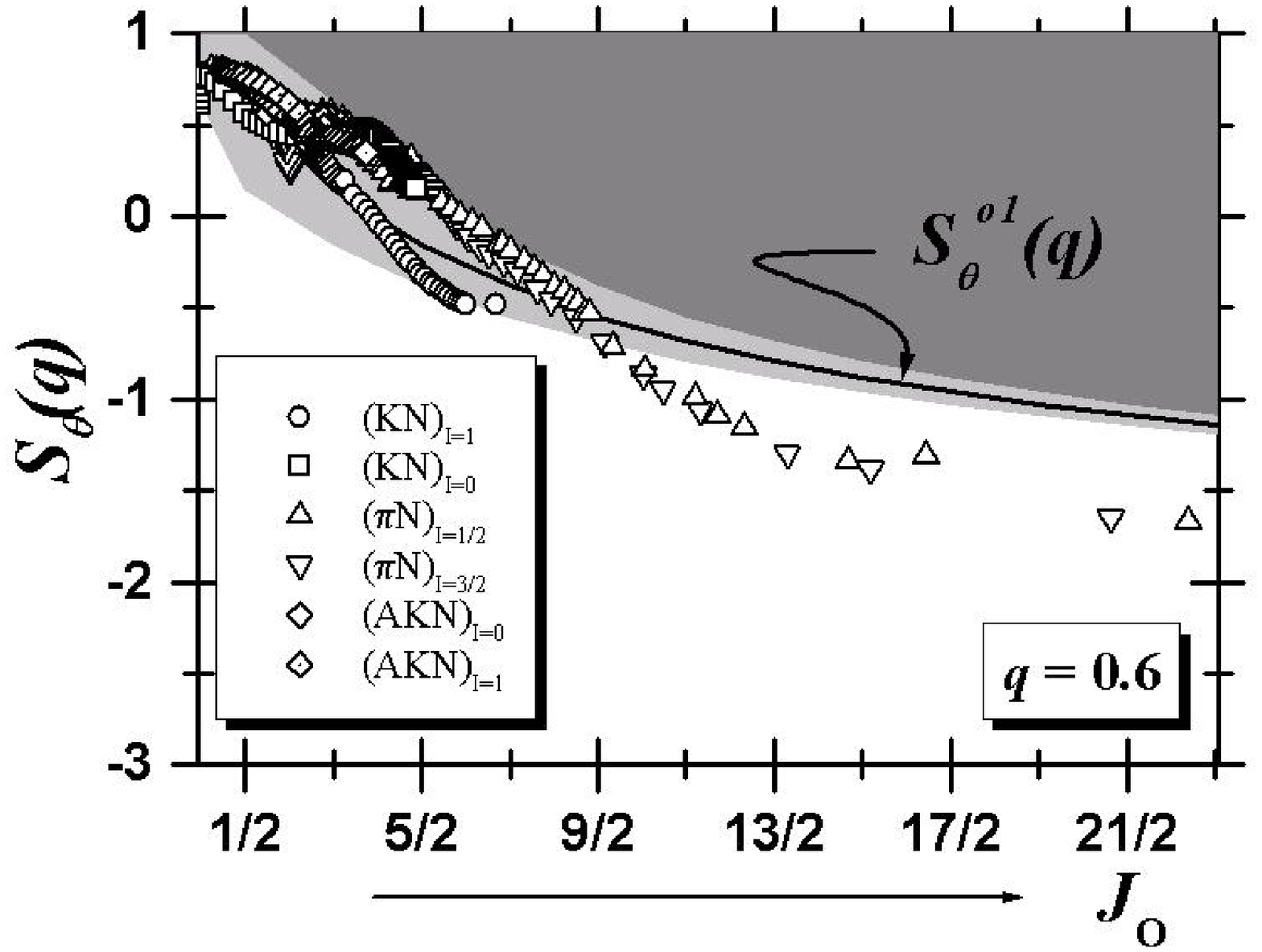}\\
\end{center}
{\bf Fig. 2:}  The experimental values of the Tsallis-like entropies $%
S_{\theta }(q)$ for $[(\pi N)_{I=1/2,3/2}$; $(KN)_{I=0,1}$; $(\overline{K}%
N)_{I=0,1}]-$ scatterings, obtained from the available experimental
phase-shifts [22-24], are compared with the PMD-SQS-optimal state
predictions $S_{\theta }^{o1}(q)$ given in Table 1 (full curve). The
saturation of the PMD-SQS (MaxEnt) optimal limits is evident.

\newpage

\noindent
\begin{center}
\epsfig{width = 90mm, file=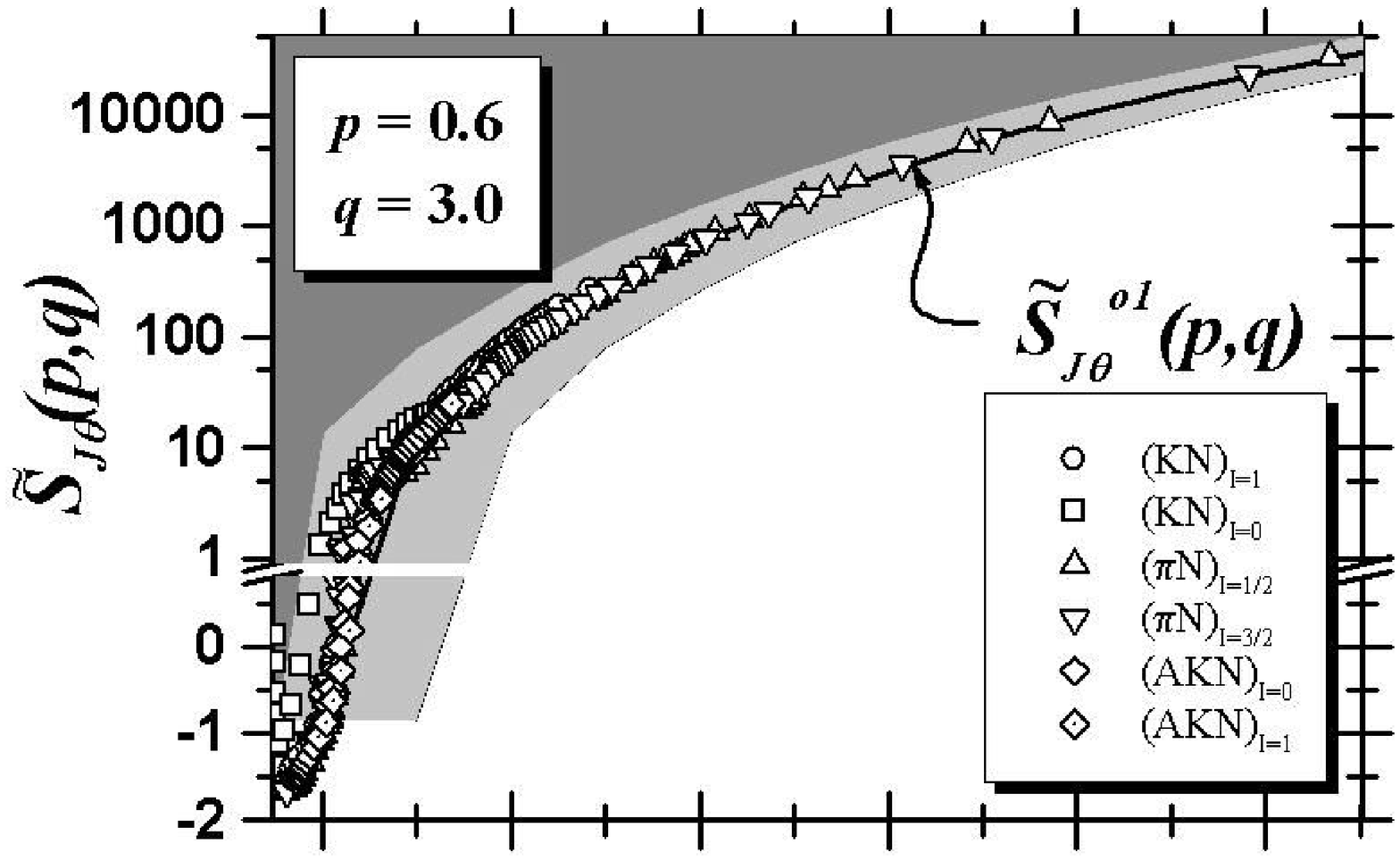}\\
\epsfig{width = 90mm, file=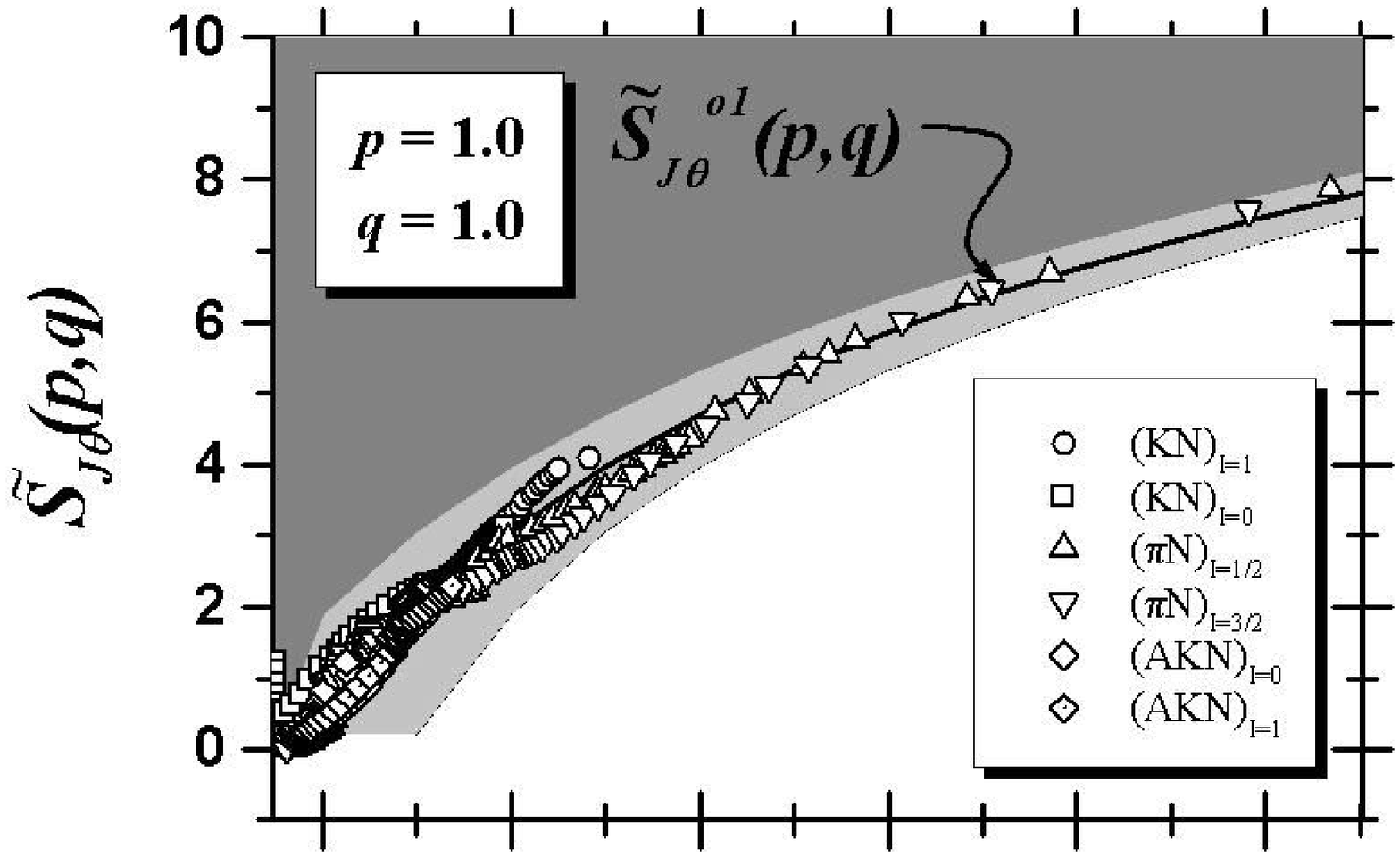}\\
\epsfig{width = 90mm, file=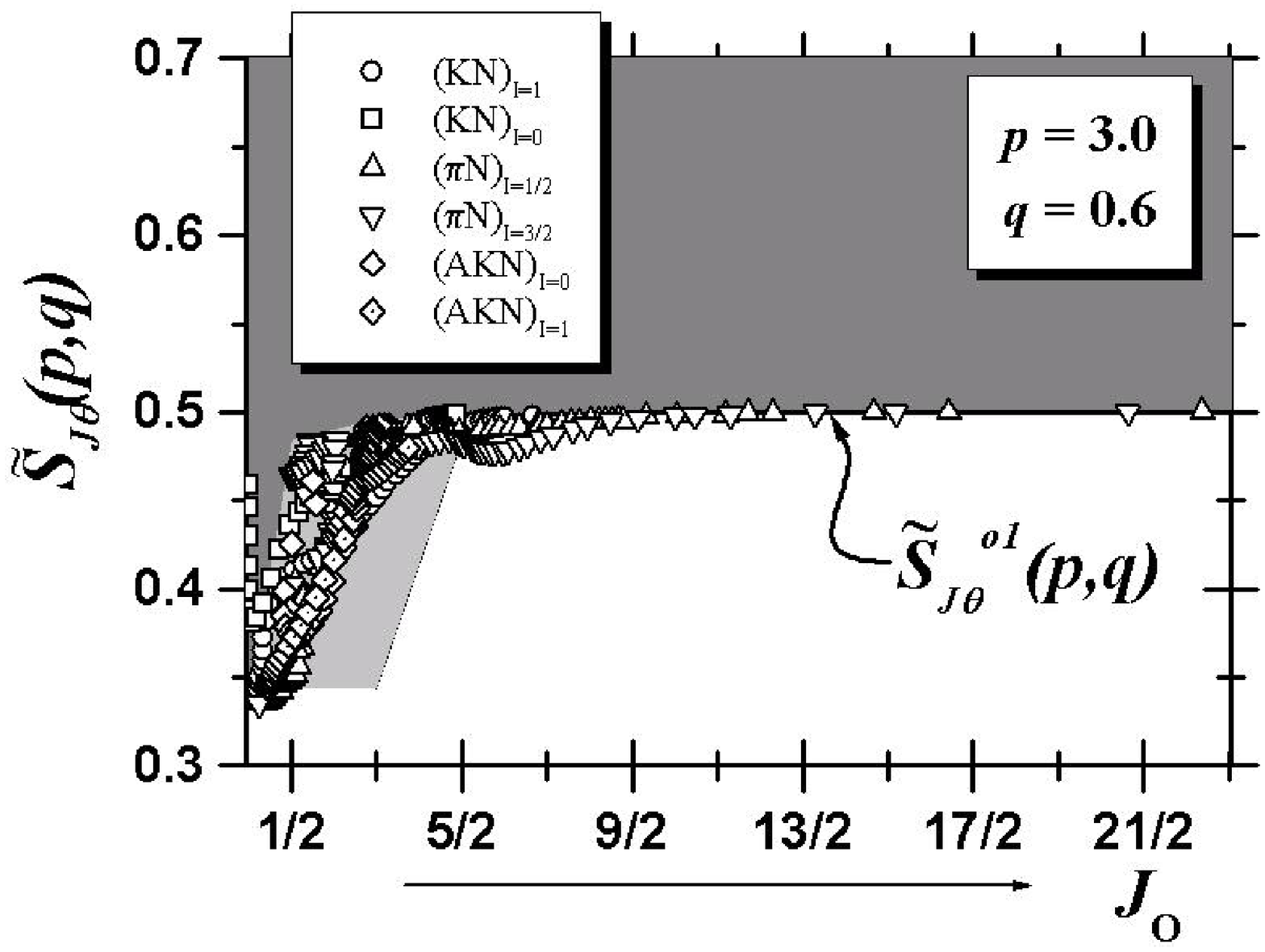}\\
\end{center}
{\bf Fig 3:} The experimental values of the Tsallis-like entropies $%
S_{J\theta }(p)$ for $[(\pi N)_{I=1/2,3/2}$; $(KN)_{I=0,1}$; $(\overline{K}%
N)_{I=0,1}]-$ scattering, obtained from the available experimental
phase-shifts [22-24], are compared with the PMD-SQS-optimal state
predictions $S_{J\theta }^{o1}(p)$ given in Table 1 (full curve). The
saturation of the PMD-SQS (MaxEnt) optimal limits is evident.


\begin{thebibliography}{99}
\bibitem{}  S. Abe, Y. Okamoto, {\it Nonextensive statistical mecanics and
its applications, }Series Lectures Notes in Physics, Berlin,
Springer-Verlag, 2001.

\bibitem{}  C. Tsallis, J. Stat. Phys. {\bf 52}, 479 (1988).

\bibitem{}  C. Tsallis and E. P. Borges, {\it Nonextensive statistical
mecanics-Applications to nuclear and high energy physics, }{\bf ArXiv}{\it :
}cond-mat/0301521 (2003), to appear in the Proceedings of the {\it X
International Workshop on Multiparticle Production Correlations and
Fluctuations in QCD} (8-15 June 2002, Crete), ed. N. Antoniou (World
Scientific, 2003); An updated bibliography on Nonextensive Statistics can be
found at the website: {\it http://tsallis.cat.cbpf.br/biblio.htm}.

\bibitem{}  I. Bediaga, E. M. F. Curado and J. Miranda, Physica {\bf A 286},
156 (2000).

\bibitem{}  C. Beck, Physica {\bf A 286}, 164 (2000). See also C. Beck,
Physica {\bf D 171}, 72 (2002).

\bibitem{}  D.B. Walton and J. Rafelski, Phys. Rev. Lett. {\bf 84}, 31
(2000).

\bibitem{}  D.B. Ion and M.L.D. Ion, Phys. Rev. Lett. {\bf 81,} 5714 (1998);
M.L.D. Ion and D.B. Ion, Phys. Rev. Lett. {\bf 83}, 463 (1999).

\bibitem{}  D.B. Ion and M.L.D. Ion, Phys. Rev. {\bf E 60}, 5261(1999);
M.L.D. Ion and D.B. Ion, Phys. Lett. {\bf B 474}, 395 (2000).

\bibitem{}  M.L.D. Ion and D.B. Ion, Phys. Lett. {\bf B 482},57 (2000).

\bibitem{}  D.B. Ion and M.L.D. Ion, Phys. Lett. {\bf B 503}, 263 (2001).

\bibitem{}  D.B. Ion and M.L.D. Ion, Phys. Lett. {\bf B 519}, 63 (2001).

\bibitem{}  D.B. Ion and M.L.D. Ion, in {\it Classical and Quantum
Complexity and Nonexten-sive Thermodynamics}, eds. P. Grigolini, C. Tsallis
and B.J. West, Chaos , Solitons and Fractals {\bf 13}, Number 3, 547
(Pergamon-Elsevier, Amsterdam, 2002); D. B. Ion and M. L. D. Ion, Rom. J.
Phys. {\bf 47, }Nr. 9-10 (2002){\bf .}

\bibitem{}  C.E. Aguiar and T. Kodama, {\it Nonextensive statistics and
multiplicity distribution in hadronic collisions}, Physica {\bf A} (2003),
in press.

\bibitem{}  O.V. Utyuzh, G. Wilkand Z. Wlodarczyk, J. Phys. {\bf G 26}, L39
(2000); G. Wilk and Z. Wlodarczyk, Phys.Rev. Lett. {\bf 84}, 2770 (2000); G.
Wilk and Z. Wlodarczyk,Physica {\bf A 305}, 227 (2002); G. Wilk and Z.
Wlodarczyk, Solitons and Fractals {\bf 13}, Number 3, 547; R. Korus, St.
Mrowczynski, M. Rybczynski and Z. Wlodarczyk, Phys. Rev. {\bf C 64}, 054908
(2001); G. Wilk and Z. Wlodarczyk, Traces of nonextensivity in particle
physics due to fluctuations, ed. N. Antoniou (World Scientific, 2003), to
appear [hep-ph/0210175].

\bibitem{}  M. Coraddu, M. Lissia, G. Mezzorani and P. Quarati,
Super-Kamiokande hep neutrinobest fit: A possible signal of nonmaxwellian
solar plasma, Physica A (2003), in press[hep-ph/0212054].

\bibitem{}  G. Kaniadakis, A. Lavagno and P. Quarati, Phys. Lett.{\bf \ B 369%
}, 308 (1996); P. Quarati, A. Carbone, G. Gervino, G. Kaniadakis, A. Lavagno
and E. Miraldi, Nucl. Phys. {\bf A 621}, 345c (1997); G. Kaniadakis, A.
Lavagno and P. Quarati, Astrophysics and spacescience {\bf 258}, 145 (1998);
G. Kaniadakis, A. Lavagno, M. Lissia and P. Quarati, in Proc.5th
International Workshop on Relativistic Aspects of Nuclear Physics (Rio de
Janeiro-Brazil, 1997); eds. T. Kodama, C.E. Aguiar, S.B. Duarte, Y. Hama, G.
Odyniec and H. Strobele (World Scientific, Singapore, 1998), p. 193; M.
Coraddu, G. Kaniadakis, A.Lavagno, M. Lissia, G. Mezzorani and P. Quarti, in
Nonextensive Statistical Mechanics and Thermodynamics, eds. S.R.A. Salinas
and C. Tsallis, Braz. J. Phys. {\bf 29}, 153 (1999); A. Lavagno and P.
Quarati, Nucl. Phys. B, Proc. Suppl.{\bf \ 87}, 209 (2000); C.M. Cossu,
Neutrini solari e statistica di Tsallis, Master Thesis, Universita degli
Studi di Cagliari (2000).

\bibitem{}  D. B. Ion, Phys. Lett. {\bf B} {\bf 376}, 282 (1996).

\bibitem{}  D. B. Ion, International J. Theor. Phys. {\bf 24}, 1217 (1985);
D. B. Ion, International J. Theor. Phys.{\bf \ 25}, 1257 (1986); D. B. Ion,
Rev. Rom. Phys. {\bf 26, }15 and 25 (1981); D. B. Ion, Rev. Rom. Phys. {\bf %
36,} 251 (1991).

\bibitem{}  S. Saitoh, {\it Theory of reproducing kernels and its
applications (}Wiley, New York, 1988).

\bibitem{}  V. M. Alecseev et al., {\it Optimalinoe Upravlenie (}Nauka,
Moskow, 1979), p. 47.

\bibitem{}  A. Zigmund, {\it Trigonometric Series} (Cambridge University
Press, Cambrdge, 1968), Vol.II, pp.102-104; J. Bergh and J. Lofstrom, {\it %
Interpolation Spaces, }(Springer-Verlag, Berlin, Hedelberg, New York, 1976),
p. 2-5.

\bibitem{}  G. H\"{o}hler et al., {\it Physics Data, Handbook of
Pion-Nucleon Sattering}, 1979, {\bf Nr.12-1.}

\bibitem{}  R. A. Arndt and L. D. Roper, Phys. Rev. {\bf D31,} 2230{\bf \ }%
(1985).

\bibitem{}  M. Alston-Garnjost, R.W. Kenney, D.L. Pollard, R.R. Ross, R.D.
Tripp, H. Nicholson, M. Ferro-Luzzi, Phys. Rev. {\bf D18}, 182 (1978).

\bibitem{}  B. M. Boghosian, Phys Rev. {\bf E 53}, 4754 (1996); C.
Anteneodo, C. Tsallis, J. Mol. Liq. {\bf 71,} 255 (1997);B. M. Boghosian,
Brazilian J. Phys. {\bf 29,} 91 (1999).
\end{thebibliography}
\end{document}